\documentclass[
 aip,
 amsmath,amssymb,
 reprint
]{revtex4-2}

\usepackage{graphicx}% Include figure files
\usepackage{dcolumn}% Align table columns on decimal point
\usepackage{bm}% bold math
\usepackage[utf8]{inputenc}
\usepackage[T1]{fontenc}
\usepackage{etoolbox}

\usepackage{xcolor}
\usepackage{float}
\usepackage[colorlinks,urlcolor=blue,citecolor=blue,linkcolor=blue]{hyperref}

\makeatletter
\def\@email#1#2{%
 \endgroup
 \patchcmd{\titleblock@produce}
  {\frontmatter@RRAPformat}
  {\frontmatter@RRAPformat{\produce@RRAP{*#1\href{mailto:#2}{#2}}}\frontmatter@RRAPformat}
  {}{}
}%
\makeatother
\begin{document}

\preprint{AIP/123-QED}

\title{Zitterbewegung Effect and Quantum Geometry in Non-Hermitian Exciton-Polariton Systems}% Force line breaks with \\

\author{Y.-M. Robin Hu}
\email{yow-ming.hu@anu.edu.au}

\author{Elena A. Ostrovskaya}%

\author{Eliezer Estrecho}
\affiliation{%
Department of Quantum Science and Technology, Research School of Physics, The Australian National University, Canberra, ACT 2601 Australia
}%

\date{\today}% It is always \today, today,
             %  but any date may be explicitly specified

\begin{abstract}
In this work, we analytically derive a semi-classical equation of motion describing the zitterbewegung effects arising in the dynamics of wavepackets in non-Hermitian systems. In Hermitian non-relativistic quantum systems, the zitterbewegung effects can arise due to the spin precession and spin-orbit coupling. Interestingly, the spin dynamics in non-Hermitian systems are qualitatively different because of the effective nonlinear terms induced by the non-Hermitian part of the Hamiltonian. In this work, we show the effects from the non-Hermitian spin dynamics by generalising the description of zitterbewegung effects to non-Hermitian systems. We also uncover novel non-Hermitian correction to the group velocity, which can be expressed in terms of the non-Hermitian quantum metric tensor in the absence of out-of-plane effective field.
\end{abstract}

%\keywords{Suggested keywords}%Use showkeys class option if keyword
                              %display desired
\maketitle

%\tableofcontents

\emph{Introduction.}---
The zitterbewegung effect, first proposed by Erwin Schrodinger in 1930 \cite{schrodinger1930}, describes the trembling motion of the relativistic electrons in the Dirac equation due to interference between states with positive and negative energies \cite{dirac1928}. Fascinatingly, it was later discovered that the zitterbewegung effect can also arise from the precession of the phase of the wavefunction \cite{Hestenes1990}, and it can also occur in non-relativistic quantum systems with spin precession and spin-orbit coupling \cite{shen2005,cserti2006}. In the past three decades, the zitterbewegung effect has been realised in a wide range of systems with spin-orbit coupling, such as semiconductor quantum wells \cite{schliemann2005,schliemann2006}, graphene \cite{rusin2007,rusin2009}, ultracold atoms \cite{vaishnav2008}, trapped ions \cite{gerritsma2010,lamata2007}, photonics \cite{zhang2008,longhi2010,ye2019,lovett2023} and exciton-polariton systems \cite{polimeno2021experimental,wen2024}.

Recently, studies on the zitterbewegung effect and related trembling-like motion have been extended to open-dissipative systems \cite{wang2017,zhan2023} described by effective non-Hermitian Hamiltonians. Research in non-Hermitian physics has attracted enormous interest~\cite{bergoltz2021,ghatak2019} primarily due to emerging topological invariants \cite{leykam2017,yao2018,zhang2020} and  edge states \cite{hatano1998,lee2019,yao2018,kunst2018,yokomizo2019,borgnia2020,okuma2020}, which could arise from novel energy degeneracies in momentum space, such as exceptional points, and Fermi arcs~\cite{zhou2018,su2020,krol2022}.The spin dynamics in non-Hermitian systems exhibits qualitatively different behavior compared to their Hermitian counterparts. The non-Hermiticity induces a nonlinear term in the mean-field equation of motion and the system becomes sensitive to the initial conditions  \cite{graefe2008,graefe2010,graefe2010iop,dattoli1990,pi2024,cius2022,zheng2021,wu2021}. This nonlinear term emerging from non-Hermiticity is expected to strongly affect the wavepacket dynamics, but to date, its role is yet to be explored. 

Quantum geometry, and in particular the quantum metric, was also recently shown to play an important role in describing the zitterbewegung effect in Hermitian two-band systems~\cite{leblanc2021}. The quantum metric is part of a larger quantity called the quantum geometric tensor (QGT)~\cite{provost1980}, which also includes the Berry curvature. The QGT plays an important role in wavepacket dynamics. In particular, the Berry curvature acts as an effective magnetic field in momentum space, resulting in an anomalous velocity transverse to the applied force~\cite{karplus1954,sundaram1999,chang1995,culcer2005}, while the quantum metric tensor plays an important role as a higher-order correction to the anomalous velocity \cite{gao2014,gao2015,bleu2018wp}, in nonlinear currents \cite{das2023,wang2023}, flat-band superconductivity \cite{chen2024,tian2023,peotta2015}, and optical second harmonic generation \cite{bhalla2022}.

More recently, there has been great interest in generalizing the QGT to non-Hermitian systems~\cite{hu2024generalized,zhang2019,ye2024,hu2025}, which is a non-trivial task due to the distinct left and right eigenvectors of the non-Hermitian Hamiltonian. There are at least two different methods of defining the QGT components including the Berry connection, within the so-called right-right (RR) and left-right (LR) formalisms. Previous works showed that the RR quantum metric plays the dominant role in wavepacket dynamics~\cite{solnyshkov2021} and that both the RR quantum metric and Berry curvature contribute to the anomalous velocity when the external force is complex valued \cite{alon2024}. The critical role of the left eigenvector was elucidated in other works showing that the LR quantum metric describes the superfluid weight in non-Hermitian flat-band superconductors \cite{he2021} and nonlinear Hall current in non-Hermitian systems \cite{qin2025}. Our recent work~\cite{hu2024generalized, hu2025} explicitly showed that both LR and RR components of the QGT are needed to accurately describe higher-order perturbative corrections to the anomalous velocity. %, and proposed an experimental protocol for measuring the LR components~\cite{hu2024generalized}.

In this work, we investigate the effect of the emergent nonlinear spin dynamics in non-Hermitian systems on wavepacket dynamics, and uncover the deep connection between the zitterbewegung effect and quantum geometry. We derive an analytical theory explicitly describing how the pseudospin dynamics manifests in the wavepacket dynamics in a two-band non-Hermitian system and demonstrate the sensitivity to initial wavepacket momentum with respect to the exceptional points and Fermi arcs. Surprisingly, in the case where the complex-valued effective magnetic field has no out-of-plane components, the non-Hermitian corrections can be explicitly described by the LR quantum metric. To demonstrate the interesting zitterbewegung effects numerically, we employ a non-Hermitian model describing exciton polaritons, hybrid light-matter quasiparticles arising from the strong coupling of excitons and confined photons in a planar microcavity \cite{deng2010,carusotto2013}. It is an accessible experimental platform for non-Hermitian physics since one can directly measure the energy, pseudospin (polarization), and density distributions through the photons emitted by polaritons when they decay \cite{gao2015,su2020}. Furthermore, the intrinsic photonic dissipation in exciton-polariton systems also allows us to study how the non-Hermiticity affects the zitterbewegung effect \cite{gao2015,su2020,krol2022}.

\emph{Theoretical Models.}---
In this work, we consider two-band non-Hermitian models in the general form 
\begin{equation}\label{eq: nh H}
    \begin{split}
        \hat{H}(\mathbf{k})&=H_0(\mathbf{k})\mathbf{I}+{\mathbf{H}}(\mathbf{k})\cdot{\bm{\sigma}}\\
        {\mathbf{H}}(\mathbf{k})&=[H_x(\mathbf{k}),H_y(\mathbf{k}),H_z(\mathbf{k})]\\
        {\bm{\sigma}}&=[\sigma_x,\sigma_y,\sigma_z]
    \end{split}
\end{equation}
where $\mathbf{I}$ is the 2$\times$2 identity matrix, $\sigma_i$ are the 2$\times$2 Pauli matrices, and ${\mathbf{H}}$ the momentum-dependent complex-valued effective magnetic field acting on the pseudospins. We split the Hamiltonian into the Hermitian part $\hat{G}$ and the non-Hermitian part $\hat{\Gamma}$ such that
    $\hat{H}=\hat{G}+i\hat{\Gamma}$, 
and separated the real and imaginary parts of the effective fields as ${\mathbf{H}}={\mathbf{G}} + i{\bm{\Gamma}}$.

The two eigenenergies take the form of $E_\pm=H_0\pm E$ with $E=\sqrt{H_x^2+H_y^2+H_z^2}$.

\begin{figure}[t]
    \centering
    \includegraphics[width=\linewidth]{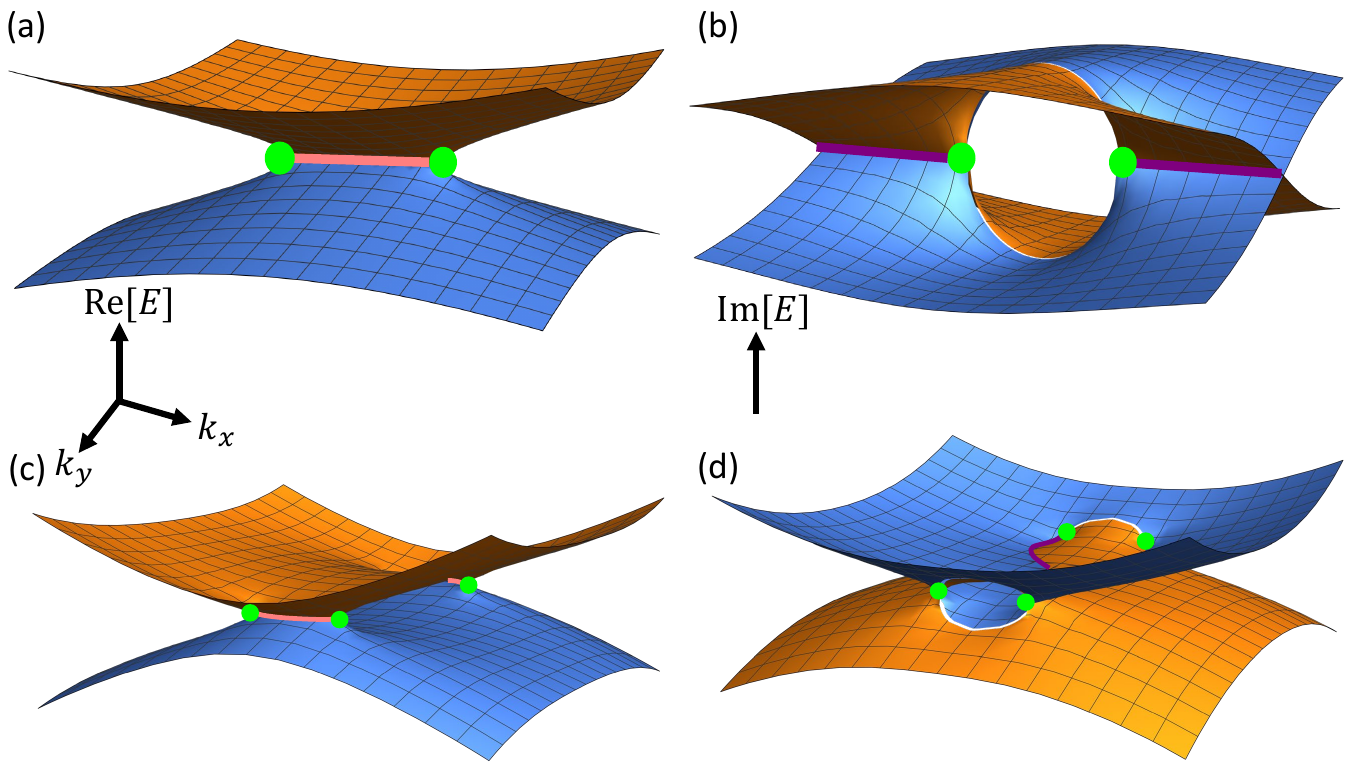}
    \caption{Real and imaginary parts of band structures of (a, b) the non-Hermitian Dirac model and (c, d) the exciton-polariton model used in this work. The green dots denote the exceptional points, and the pink and purple lines highlight the bulk and imaginary Fermi arcs, respectively.}
    \label{fig: band}
\end{figure}

The first model we consider is the non-Hermitian Dirac model described by~\cite{solnyshkov2021}
\begin{equation}\label{eq: nh dirac}
    \hat{H}(\mathbf{k})=k_x\sigma_x+(k_y-i\kappa)\sigma_y.
\end{equation}
In the Hermitian case ($\kappa=0$), the model has a degeneracy known as the Dirac point at $\mathbf{k}=(0,0)$. When $\kappa \neq 0$, the Diract point splits into a pair of exceptional points at $\mathbf{k}=(\pm\kappa,0)$. The pair is connected by the bulk Fermi arc (where $\operatorname{Re}[E]=0$) and the imaginary Fermi arc (where $\operatorname{Im}[E]=0$) \cite{zhou2018,su2020}. In this model, both Fermi arcs lie along the $k_x$-axis in momentum space, with the bulk Fermi arc on $|k_x|<|\kappa|$ while the imaginary Fermi arcs on $|k_x|>|\kappa|$, extending to infinity [Fig.~\ref{fig: band}(a,b)].

The second model we consider is based on a perovskite-based exciton-polariton system\cite{su2020} with 
\begin{equation}\label{eq: nh perovskite}
    \begin{split}
        H_0(\mathbf{k})&=E_0-i\gamma(\mathbf{k})+\frac{\hbar^2\mathbf{k}^2}{2m}\\
        H_x(\mathbf{k})&=(\alpha-ia)+(\beta-ib)(k_x^2-k_y^2)\\
        H_y(\mathbf{k})&=2(\beta-ib)k_xk_y\\
        H_z(\mathbf{k})&=\Delta
    \end{split}
\end{equation}
where $E_0,\gamma$ are the mean energy and linewidth, $m$ is the effective exciton-polariton mass, $\alpha$ is the cavity anisotropy that splits the linearly-polarized modes, $\beta$ is the TE-TM splitting arising from photonic spin-orbit coupling, and $\Delta$ is the Zeeman splitting arising from an external magnetic field. Here, $a,b$ represent the linewidth splitting due to cavity anisotropy and TE-TM splitting. The momentum dependence of the mean-polariton linewidth keeps the imaginary parts of both eigenenergies strictly negative, since the polariton system we consider is dissipative, but this term is kept general since it does not play an important role in the dynamics~\cite{hu2023wave}. When $\Delta=0$, there are two pairs of exceptional points in momentum space. Each pair is connected by the Fermi arcs as shown in Fig.~\ref{fig: band}(c, d). A large enough $\Delta$ will annihilate each pair of exceptional points and open an energy gap~\cite{su2020,hu2023wave}.

In the following section, we will demonstrate the dynamics in the pseudospins and the wavepackets using the exciton-polariton model. However, we also keep the main results general so it can be used to describe any non-Hermitian two-band system described by Eq.~(\ref{eq: nh H}). Note that the models considered here have three-dimensional orientation of the pseudospin and effective magnetic field, but the motional degrees of freedom are constrained on the $x$-$y$ plane.

\begin{figure}[t]
    \centering
    \includegraphics[width=0.45\textwidth]{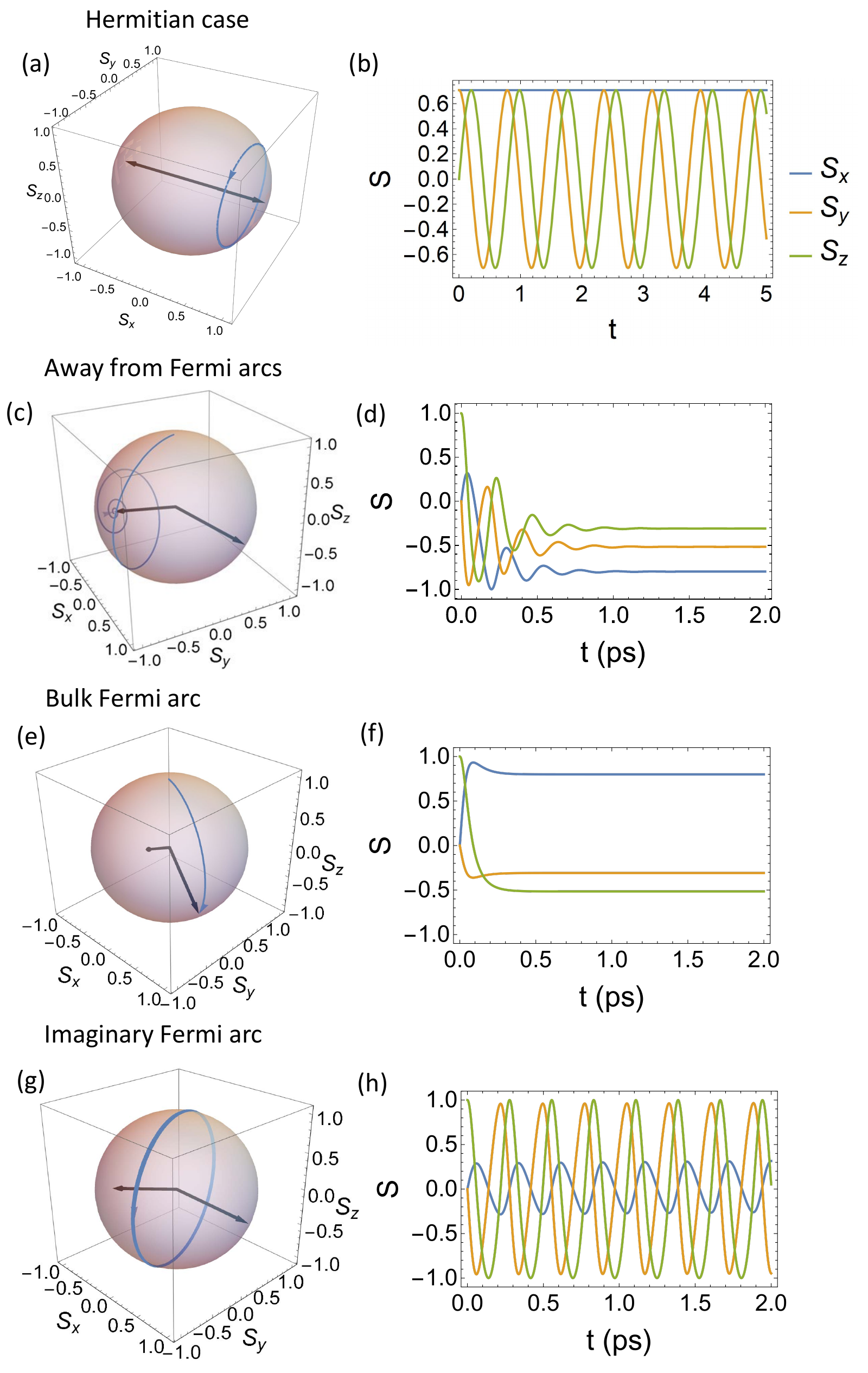}
    \caption{Pseudospin dynamics for a (a,b) Hermitian and (c-h) non-Hermitian  in an exemplar Hermitian system plotted in different components in (a) in on the Bloch sphere in (b). Here, the blue arrow denotes the trajectory of the pseudospin and the black arrows denote the pseudospins of the two eigenstates. Also showing showing the pseudospin dynamics. (c,d) are away from the Fermi arcs at $\mathbf{k}=(1.5,1.5)\mu$m$^{-1}$, (e,f) are on the bulk Fermi arc at $\mathbf{k}=(0.5,2.685)\mu$m$^{-1}$ and (g,h) are on the imaginary Fermi arc at $\mathbf{k}=(0.9,1.22)\mu$m$^{-1}$. The initial pseudospin configuration is chosen to be $\mathbf{S}=(1/\sqrt{2},1/\sqrt{2},0)$ and $\mathbf{S}=(0,0,1)$ for the Hermitian and non-Hermitian cases, respectively.}
    \label{fig: new s}
\end{figure}

\emph{Pseudospin Dynamics.}---
In Hermitian systems ($\hat{\Gamma}=0$), the effective magnetic field ${\mathbf{G}}$ generates the precession of the pseudospins at frequency $\omega = 2E/\hbar$, where $E=\sqrt{{\mathbf{G}}\cdot{\mathbf{G}}}$ is the mean-subtracted eigenenergy [see Fig.~\ref{fig: new s}]. This can be seen from the equation of motion that describes the pseudospin ${\mathbf{S}}$ dynamics in the Hermitian limit,
\begin{equation}
    \frac{d}{dt}{\mathbf{S}}=\frac{2}{\hbar}\Big({\mathbf{S}}\times{\mathbf{G}}\Big),
\end{equation}
which has a general solution of the form of
\begin{equation}\label{eq: hermitian s dynmics}
    {\mathbf{S}}(t)=\frac{{\mathbf{G}}\cdot{\mathbf{S}_0}}{E^2}{\mathbf{G}}-\frac{{\mathbf{G}}\times({\mathbf{G}}\times{\mathbf{S}_0})}{E^2}\cos{\omega t}-\frac{{\mathbf{G}}\times{\mathbf{S}}_0}{E}\sin{\omega t}
\end{equation}
with initial condition ${\mathbf{S}}(t=0)={\mathbf{S}}_0$.

When the system is non-Hermitian, the imaginary part of the effective magnetic field $\bm{\Gamma}$ gives rise to additional terms in the equation of motion and the dynamics of the pseudospins can be described by the equation~\cite{graefe2008}
\begin{equation}\label{eq: nh S}
    \frac{d}{dt}{\mathbf{S}}=\frac{2}{\hbar}\left[{\mathbf{S}}\times{\mathbf{G}}-({\bm{\Gamma}}\cdot{\mathbf{S}}){\mathbf{S}}+{\bm{\Gamma}}\right]
\end{equation}
with the additional nonlinear term of $O(S^2)$ arising from non-Hermiticity \cite{graefe2008,graefe2010,graefe2010iop,pi2024,cius2022,zheng2021,wu2021}. This equation closely resembles the Landau-Lifshitz-Gilbert equation describing a damped spin precession \cite{gilbert2004,landau1935},
\begin{equation}\label{eq: llg equation}
       \frac{d}{dt}{\mathbf{S}}=\frac{2}{\hbar}\left[{\mathbf{S}}\times{\mathbf{G}}-(\lambda{\mathbf{G}}\cdot{\mathbf{S}}){\mathbf{S}}+\lambda{\mathbf{G}}\right],
\end{equation}
but the damping term $\lambda{\mathbf{G}}$ (where $\lambda$ is some damping constant ) is replaced by another imaginary-valued effective field ${\bm{\Gamma}}$. Note that despite the equations of motion in Eq.~(\ref{eq: nh S}, \ref{eq: llg equation}) describe dissipative systems, the components in both equations are real-valued, since the pseudospins here are defined and normalized in the RR formalism \cite{graefe2008}~.

To solve the non-Hermitian pseudospin dynamics, we calculate $\mathbf{S}(t)=[S_x,S_y,S_z]$ using the time-evolution operator
\begin{equation}
    S_j=\frac{\langle\psi_0|e^{i\hat{H}^\dagger t/\hbar}\sigma_j e^{-i\hat{H}t/\hbar}|\psi_0\rangle}{\langle\psi_0|e^{i\hat{H}^\dagger t/\hbar}e^{-i\hat{H}t/\hbar}|\psi_0\rangle}
\end{equation}
where $|\psi_0\rangle$ denotes the initial spinor wavefunction. The time evolution operator can be defined using the usual form of matrix exponential in terms of the components of the effective magnetic field $H_j$ \cite{dattoli1990}
\begin{equation}
    e^{-\frac{i\hat{H}t}{\hbar}}=e^{-\frac{iH_0 t}{\hbar}}\mathbf{I}\cdot\begin{pmatrix}
        \cos{\frac{Et}{\hbar}}-\frac{iH_z}{E}\sin{\frac{Et}{\hbar}} && -i\frac{H_x-iH_y}{E}\sin{\frac{Et}{\hbar}} \\ -i\frac{H_x+iH_y}{E}\sin{\frac{Et}{\hbar}} &&  \cos{\frac{Et}{\hbar}}+\frac{iH_z}{E}\sin{\frac{Et}{\hbar}}
    \end{pmatrix}
\end{equation}
which is valid except at the exceptional points, where the two eigenstates coalesce and the Hamiltonian operator is no longer diagonalizable \cite{shen2018}. At the exceptional points, the dynamics can be determined by solving the differential equation in Eq.~(\ref{eq: nh S}).

As an illustrative example, we consider a simple case of an in-plane effective magnetic field, i.e. $G_z=\Gamma_z=0$, and the initial pseudospin configuration $\mathbf{S}_0=[0,0,1]$. In this case, the pseudospins dynamics is
\begin{equation}\label{eq: S0Sample}
    \begin{split}
    S_x=&\frac{2\operatorname{Im}\Big[\sqrt{H_+H_-^*}\Big]\sin{\omega_r t}+2\operatorname{Re}\Big[\sqrt{H_+H_-^*}\Big]\sinh{\omega_i t}}{(|H_-|-|H_+|)\cos{\omega_r t}+(|H_-|+|H_+|)\cosh{\omega_i t}}\\
        S_y=& \frac{-2\operatorname{Re}\Big[\sqrt{H_+H_-^*}\Big]\sin{\omega_r t}-2\operatorname{Im}\Big[\sqrt{H_+H_-^*}\Big]\sinh{\omega_i t}}{(|H_-|-|H_+|)\cos{\omega_r t}+(|H_-|+|H_+|)\cosh{\omega_i t}}\\
        S_z=& \frac{(|H_-|+|H_+|)\cos{\omega_r t}+(|H_-|-|H_+|)\cosh{\omega_i t}}{(|H_-|-|H_+|)\cos{\omega_r t}+(|H_-|+|H_+|)\cosh{\omega_i t}},
    \end{split}
\end{equation}
where the complex frequency is $\omega_r + i\omega_i = 2E/\hbar$. We denote $H_\pm=H_x\pm iH_y$ to simplify the expressions. Similar to the Hermitian case, the pseudospins oscillate but with the frequency of $\omega_r = 2\operatorname{Re}E/\hbar$. However, there are two new effects: (i) the oscillations are no longer harmonic due to the denominators and (ii) the amplitude of oscillation is decaying due to the $\omega_i = 2\operatorname{Im}E/\hbar$.

We emphasize that although Eq.~(\ref{eq: S0Sample}) is not a simple closed form solution like in the Hermitian case [see Eq.~(\ref{eq: hermitian s dynmics})], the non-Hermitian pseudospin dynamics presented here, in general, is always analytic and deterministic. We chose the simple case with $H_z=0$ for simplicity, and the behaviors are not qualitatively different when $H_z\neq 0$, as the latter only tilts the effective field.

The pseudospin dynamics features distinct behavior in special cases, such as at the exceptional point or on the Fermi arcs. When away from both the imaginary and the bulk Fermi arcs, the pseudospin will oscillate while decaying towards the less dissipative eigenstate  [see Figs.~\ref{fig: new s}(c,d)].  While on the bulk Fermi arc, there is no oscillation, and the pseudospin will align with the pseudospin of the less dissipative eigenstate  [see Figs.~\ref{fig: new s}(e,f)]. Similarly, on the exceptional points, where the two eigenstates coalesce, the pseudospin also aligns with the eigenstate with no oscillation. At the imaginary Fermi arc, where $\omega_i=0$, there is no decay in the amplitude of the pseudospin precession, similar to the Hermitian case, but the oscillation is anharmonic [see Figs.~\ref{fig: new s}(g,h)].

\emph{Results and Discussion: Zitterbewegung Effects.}---
The zitterbewegung effects can be realized in the two-band system by initiating the wavepacket in a superposition of two eigenstates. In the Hermitian case, the dynamics of the wavepacket center of mass (COM) position $\mathbf{r}_{H}$ can be described by the equation of motion~\cite{sedov2018}
\begin{equation}
\hbar\Dot{\mathbf{r}}_{H,i}=\left.\partial_{k_i}G_0+\partial_{k_i}\mathbf{G}\cdot\mathbf{S}\right|_{\mathbf{k}=\mathbf{k}_c},
\end{equation}
where $\mathbf{k}_c$ denotes the COM momentum of the wavepacket. The pseudospin precession and non-vanishing $\partial_\mathbf{k}\mathbf{G}$ arising from spin-orbit coupling result in the oscillation of the wavepacket centre-of-mass motion \cite{sedov2018}, which is the zitterbewegung effect.

In the non-Hermitian case, the dynamics is much richer and the corresponding COM equation of motion is:
\begin{equation}\label{eq: ZBW}
    \begin{split}
        \hbar\Dot{\mathbf{r}}_i=&\hbar\Dot{\mathbf{r}}_{H,i}+\frac{2t}{\hbar}
        \bm{\Gamma}\cdot \left[\partial_{k_i}\mathbf{G} - \partial_{k_i}\bm{\Gamma}\times\mathbf{S} - \mathbf{S}(\partial_{k_i}\mathbf{G}\cdot\mathbf{S})\right]
        \\
        &+\frac{2t}{\hbar}\bm{\Gamma}\cdot \left[ \operatorname{Re}[\mathbf{B}_i] - \operatorname{Im}[\mathbf{B}_i]\times\mathbf{S} - \mathbf{S}\left(\operatorname{Re}[\mathbf{B}_i]\cdot\mathbf{S}\right)\right] 
    \end{split}
\end{equation}
where $\mathbf{B}_i$ is a notation to simplify the equation of motion and is defined as
\begin{equation}
\begin{split}
    \mathbf{B}_i=&-\frac{1}{2\hbar\omega^2t}\left(\cos\omega t-1\right)(\mathbf{H}\times\partial_{k_i}\mathbf{H})\\
    &-\frac{1}{2\hbar^2\omega^2}\left(\frac{\sin{\omega t}}{\omega t}-1\right)(\mathbf{H}\times(\mathbf{H}\times\partial_{k_i}\mathbf{H})).
\end{split}
\end{equation}
In this form, $\mathbf{B}_i$ plays the same role as the $k_i$-derivatives of the effective magnetic field $\mathbf{H}$. Note that although $\mathbf{B}_i$ can be non-zero in Hermitian systems, all of the corrections in terms of $\mathbf{B}_i$ are proportional to $\bm{\Gamma}$, and hence will vanish in the Hermitian limit.

Interestingly, for the common case where there is no out-of-plane field, i.e. $H_z=0$, the non-Hermitian correction can be simplified to
\begin{equation}\label{eq: Hz=0 ZBW}
\begin{split}
    \mathbf{B}_i=\sqrt{g_{k_ik_i}^{LR}}\left[\left( \frac{\sin{\omega t}}{\omega t}-1\right)\Big(H_x\hat{\mathbf{y}}-H_y\hat{\mathbf{x}}\Big)
    -\frac{\hbar}{t}\left(\cos\omega t-1\right)\hat{\mathbf{z}} \right]
\end{split}
\end{equation}
where $g_{\mathbf{kk}}^{LR}=(g_{k_xk_x}^{LR},g_{k_yk_y}^{LR})$ represents the diagonal elements of the left-right quantum metric tensor (LR QMT)~\cite{hu2024generalized}. It satisfies the identity $g_+^{LR}=g_-^{LR}=g^{LR}$ in a two-band system. The diagonal components of LR QMT, in the $H_z=0$ case, take the form
\begin{equation}
g_\mathbf{kk}^{LR}=\frac{(H_x\partial_\mathbf{k}H_y-H_y\partial_\mathbf{k}H_x)^2}{4E^4}.
\end{equation}
This is a key finding of this work, explicitly showing the role of the quantum metric tensor on the zitterbewegung effect. For more details of the derivation of Eqs.~(\ref{eq: ZBW}, \ref{eq: Hz=0 ZBW}), please refer to the Supplemental Material. 

\begin{figure}[t]
    \centering
    \includegraphics[width=\linewidth]{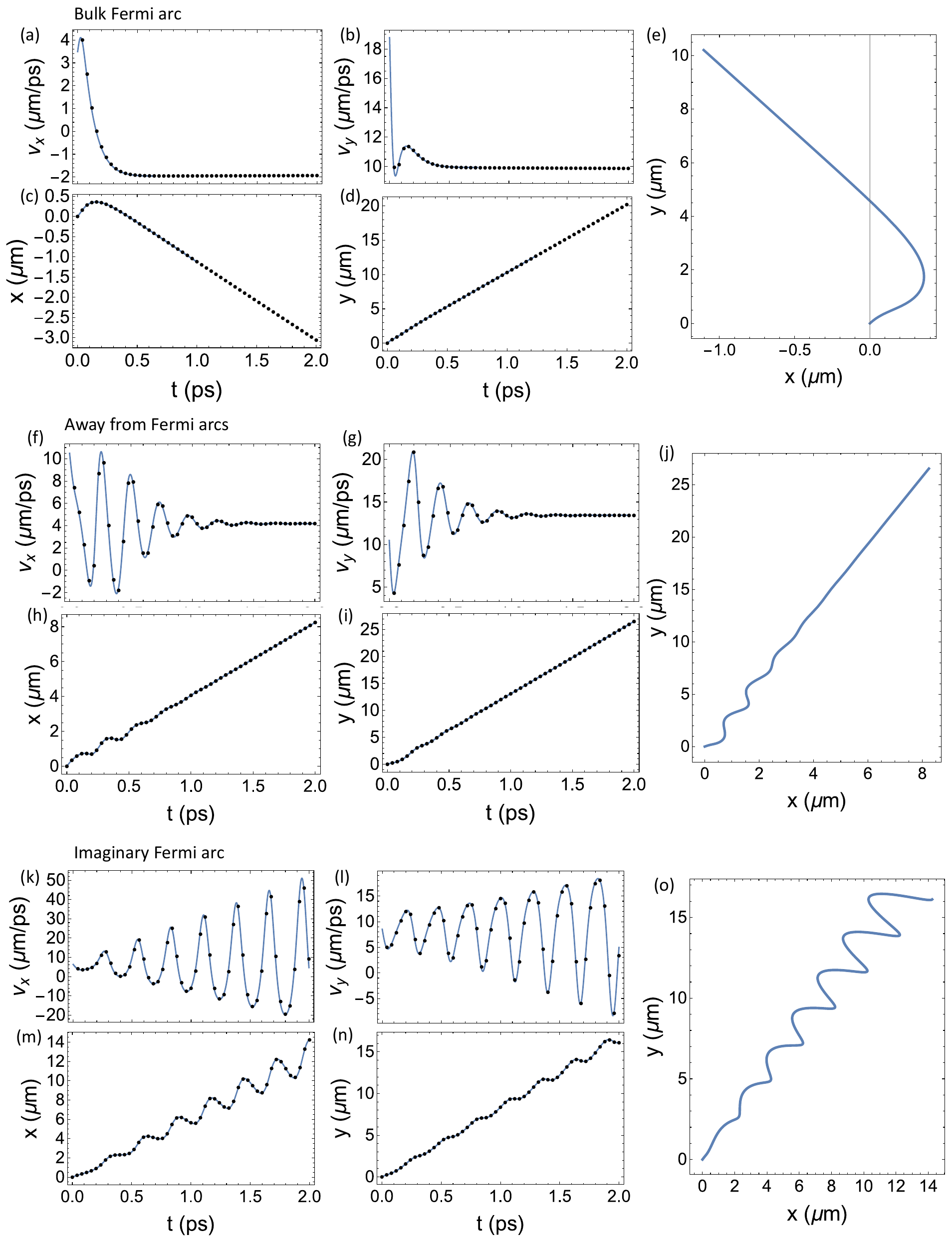}
    \caption{The group velocities (a,b,f,g,k,l) and center-of-mass positions (c,d,h,i,m,n) of a wavepacket initially centred at (a-d) $\mathbf{k}=(0.5,2.685)~\mu$m$^{-1}$ (on the bulk Fermi arc),  (f-i) $\mathbf{k}=(1.5,1.5)~\mu$m$^{-1}$ (away from the Fermi arcs) and (k-n) $\mathbf{k}=(0.9,1.22)~\mu$m$^{-1}$ (on the imaginary Fermi arc) as a function of time. Also showing the trajectories in real space in (e,j,o). The lines are analytic results derived from Eq.~(\ref{eq: Hz=0 ZBW}) and the black dots are the numerical results for a wavepacket with initial with of $0.005~\mu$m$^{-1}$ in momentum space.}
    \label{fig: polariton zbw}
\end{figure}

We emphasize that in the derivation of Eqs.~(\ref{eq: ZBW}, \ref{eq: Hz=0 ZBW}), we assume that the initial pseudospin configuration is homogeneous in momentum space, and the wavepackets are infinitesimally small in momentum space to rule out self-acceleration \cite{hu2023wave}. %Self-acceleration becomes relevant when the width of the wavepacket is large in momentum space such that different momentum components undergoes significantly different decay rates. Furthermore, the large wavepacket can also split in certain conditions (like when initialized on the imaginary Fermi arc)~\cite{hu2023wave}, resulting in a more complicated dynamics and the breakdown of the semi-classical approximation.

To provide a concrete example of the dynamics, we simulated the dynamics of a wavepacket that is narrow but with a finite width in momentum space using the exciton-polariton model described by the Hamiltonian in Eq.~(\ref{eq: nh perovskite}). The wavepackets is initialized at different momenta to demonstrate the effects described in Eq.~(\ref{eq: Hz=0 ZBW}).

When the wavepacket is placed on the bulk Fermi arc, the group velocity will decay to a constant with no oscillation [see Fig.~\ref{fig: polariton zbw} (a-e)]. This agrees with the overdamped psudospin precession shown in Fig.~\ref{fig: new s}(e,f).
When the wavepacket is initialized away from the Fermi arcs, the group velocity will oscillate anharmonically with decreasing amplitude, and eventually decay to a constant group velocity [see Fig.~\ref{fig: polariton zbw} (f-j)]. Again, this is a similar behaviour to the pseudospin precession presented in Fig.~\ref{fig: new s}(c,d). When the wavepacket is initialized on the imaginary Fermi arc, it will oscillate anharmonically in time, similar to the behaviour of the pseudospin precession in Fig.~\ref{fig: new s}(g,h). However, the amplitude of oscillation will grow with time, as shown in Fig.~\ref{fig: polariton zbw} (k-o). Linear growth is due to the terms proportional to $2t/\hbar^2$ in Eq.~(\ref{eq: Hz=0 ZBW}).

Hence, in a non-Hermitian system, a sustained zitterbewegung effect is only possible when the wavepacket is initiated on the imaginary Fermi arc. However, the amplitude grows with time. In other initial conditions, the oscillation in group velocity decays quickly decays. 

The semi-classical equation in Eqs.~(\ref{eq: ZBW}, \ref{eq: Hz=0 ZBW}) are complicated and difficult to gain intuition from. However, for the non-Hermitian Dirac model with $\Delta=0$ described by the Hamiltonian in Eq.~(\ref{eq: nh dirac}), we can simplify the equations of motion on the bulk and imaginary Fermi arcs as an illustration. Here, we set $\hbar=1$ for simplicity.

The diagonal elements of the left-right quantum metric tensor for the non-Hermitian Dirac model take the forms \cite{hu2024generalized}
\begin{equation}
    \begin{split}
        g_{xx}^{LR}&=\frac{(k_y-i\kappa)^2}{4E^4}\\
        g_{yy}^{LR}&=\frac{k_x^2}{4E^4}.
    \end{split}
\end{equation}
Note that $\sqrt{g_{xx}^{LR}}$ is purely imaginary and $\sqrt{g_{yy}^{LR}}$ is purely real on both the bulk and imaginary Fermi arcs. Using these identities, we can write down a simple form of Eq.~(\ref{eq: Hz=0 ZBW}) for the group velocity $(v_x,v_y)$. The equations of motion on the imaginary Fermi arc are:
\begin{equation}\label{eq: dirac imaginary fermi}
    \begin{split}
        v_x=&S_x+2t\kappa S_x S_y-2i\sqrt{g_{xx}^{LR}}\bigg(-(\cos{\omega_r t}-1)\kappa S_x\\
        &-2\Big(\frac{\sin{\omega_r t}}{\omega_r}-t\Big)\kappa^2 S_x S_y\bigg)\\
         v_y=&S_y-2t\kappa (1-S_y^2))+2\sqrt{g_{yy}^{LR}}\bigg(2\Big(\frac{\sin{\omega_r t}}{\omega_r}-t\Big)\\
         &\times(-\kappa k_x-\kappa^2 S_z+\kappa k_x S_y^2)-(\cos{\omega_r t}-1)\kappa S_y S_z\bigg),
    \end{split}
\end{equation}
and on the bulk Fermi arc are:
\begin{equation}\label{eq: dirac bulk fermi}
    \begin{split}
        v_x=&S_x+2t\kappa S_x S_y-2i\sqrt{g_{xx}^{LR}}\bigg(-(\cosh{\omega_it}-1)\kappa S_x\\
        &-2\Big(\frac{\sinh{\omega_it}}{\omega_i}-t\Big)\kappa^2 S_x S_y\bigg)\\
         v_y=&S_y-2t\kappa (1-S_y^2))+2\sqrt{g_{yy}^{LR}}\bigg(2\Big(\frac{\sinh{\omega_it}}{\omega_i}-t\Big)\\
         &\times(-\kappa k_x-\kappa^2 S_z+\kappa k_x S_y^2)-(\cosh{\omega_it}-1)\kappa S_y S_z\bigg).
    \end{split}
\end{equation}
The equations of motion in these two special cases highlight the oscillation resulting from $\omega_r = 2\operatorname{Re}[E]$ and decay in the amplitude of the oscillation as a direct result of $\omega_i=2\operatorname{Im}[E]$. The latter is absent when the wavepacket is on the imaginary Fermi arc. For reference, we presented the corresponding time-evolution of the pseudospin $\mathbf{S(t)}$ in the Supplemental Material.

\begin{figure}[t]
    \centering
    \includegraphics[width=\linewidth]{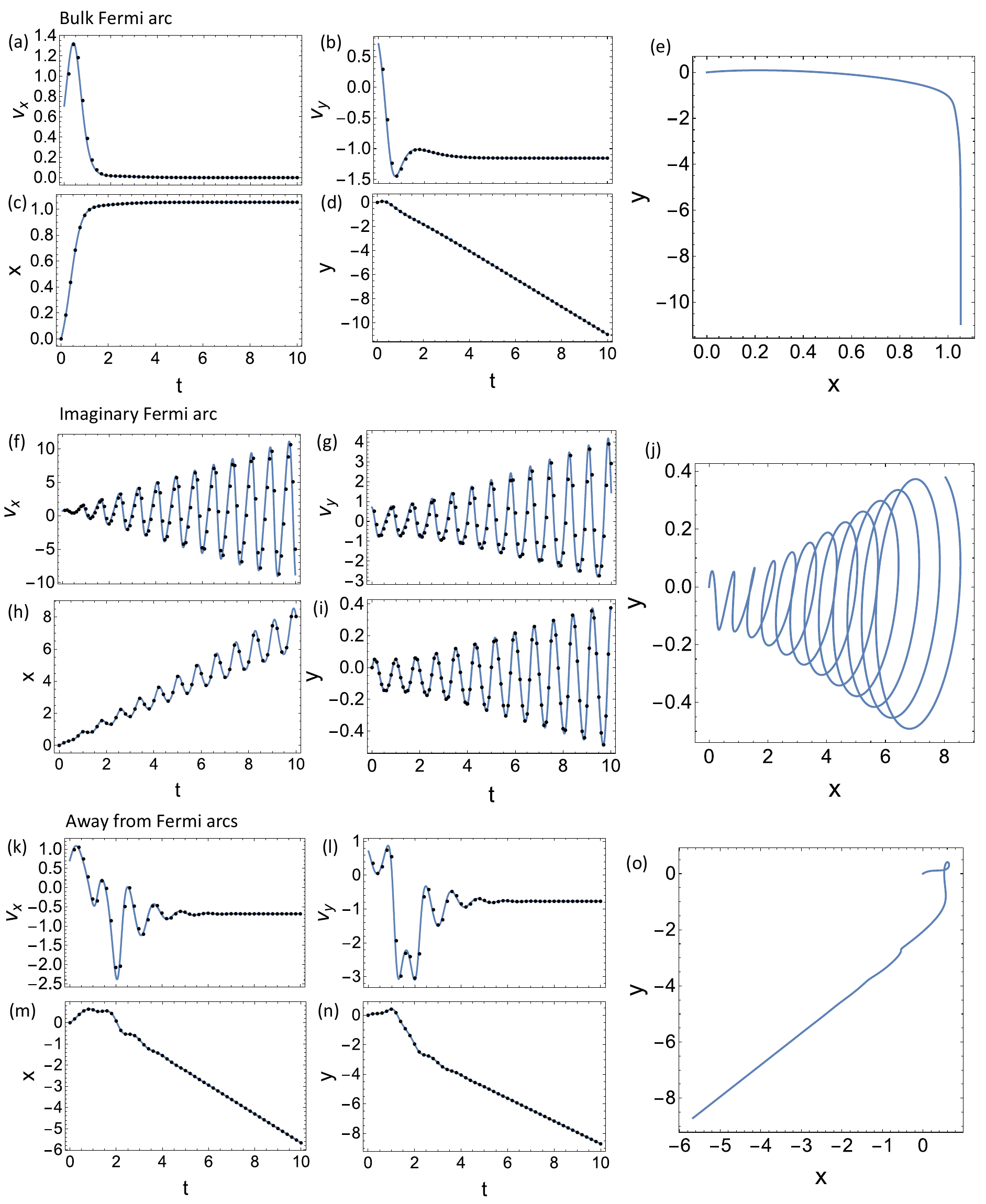}
    \caption{The group velocities (a,b,f,g,k,l) and centre-of-mass positions (c,d,h,i,m,n) of a wavepacket initially centred at (a-d) $\mathbf{k}=(0.5,0)$ (on the bulk Fermi arc),  (f-i) $\mathbf{k}=(4,0)$ (on the imaginary Fermi arc) and (k-n) $\mathbf{k}=(2,2)$ (away from the Fermi arcs) as a function of time. Also showing the trajectories in real space in (e,j,o). The lines represent analytic results derived from Eq.~(\ref{eq: Hz=0 ZBW}) and the black dots represent the numerical results for a wavepacket with initial with of $0.005$ in momentum space.}
    \label{fig: dirac zbw}
\end{figure}

Similar to the exciton polaritons in the previous Section, we also simulate wavepacket dynamics to show that it agrees with the semi-classical equation of motions in the non-Hermitian Dirac model. Here, the wavepacket is also narrow in $k$-space and has initial uniform pseudospin texture of ${\mathbf{S}}_0=(1/\sqrt{2},1/\sqrt{2},0)$. This was chosen since if the $S_x$ ($S_y$) component of the initial wavepacket is zero, $v_x$ ($v_y$) is also zero, as can be seen from Eqs.~(\ref{eq: dirac imaginary fermi}, \ref{eq: dirac bulk fermi}). Just like the case with the exciton polariton model, the simulation show strong agreement with the analytic theory as shown in Fig.~\ref{fig: dirac zbw}]. There is also strongly dependence on the initial momentum as starkly shown when the wavepacket is on the bulk or imaginary Fermi arcs or away from the Fermi arcs.

\begin{figure}[t]
    \centering
    \includegraphics[width=0.45\textwidth]{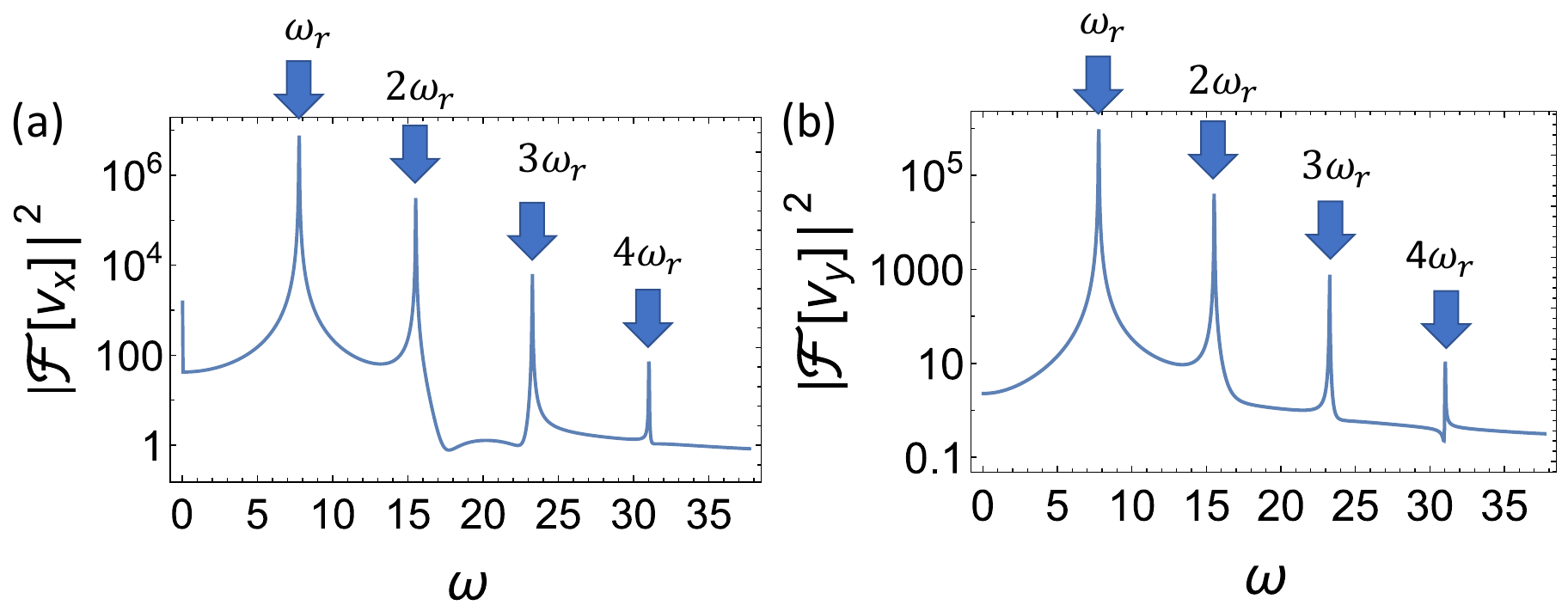}
    \caption{The Fourier transform of the group velocities $v_x$ and $v_y$ on the imaginary Fermi arc at $\mathbf{k}=(4,0)$ in the non-Hermitian Dirac model. Here, the primary frequency $\omega$ is shown on both plots along with doubled and tripled frequencies.}
    \label{fig: zbw fourier}
\end{figure}

Finally, we look closely at the anharmonicity of the zitterbewegung which arises from the emergent nonlinear terms of $O(S^2)$ in both the pseudospin dynamics~\cite{graefe2008,graefe2010,graefe2010iop,pi2024,cius2022,zheng2021,wu2021} [see Eq.~(\ref{eq: nh S})] and the COM equation of motion [see e.g., Eq.~(\ref{eq: dirac imaginary fermi})]. These terms give rise to frequency doubling or generation of higher harmonics in the oscillation, which is not seen in the usual Hermitian cases. To show this, we plot the Fourier spectra of the group velocities when the wavepacket is initialized on the imaginary Fermi arc in Fig.~\ref{fig: zbw fourier}.

The spectra show multiple harmonics, at least up to $4\omega_r$ with $\omega\approx7.6969$, $\omega\approx15.551$ and $\omega\approx23.311$ and $\omega=31.0704$, corresponding to oscillation at frequencies of $\omega_r=7.7460$, $2\omega_r=15.4919$, $3\omega_r=23.2379$ and $4\omega_r=30.9839$. These are clear evidence of anharmonic oscillation, induced by the nonlinear terms. Hence, the anharmonic oscillation is another effect of non-Hermiticity on the zitterbewegung of the wavepacket.

\emph{Conclusion.}---
In this work, we derive an equation of motion of the wavepacket centre of mass which describes a non-Hermitian generalisation of zitterbewegung effect. We uncover two novel effects on the zitterbewegung effect due to non-Hermiticity: (1) The oscillation is sensitive to the initial momentum, relative to the Fermi arcs, similar to the nonlinear dynamics of a non-Hermitian Bose-Einstein dimer \cite{graefe2008,graefe2010,graefe2010iop}. It can be overdamped (on the bulk Fermi arc), damped (away from the arcs) or growing in time (on the imaginary Fermi arc). (2) The oscillation is inherently anharmonic due to nonlinear terms in the equations describing both the pseudospin and wavepacket COM dynamics.

We demonstrated strong agreement of our analytical theory with numerical simulations of wavepacket dynamics using a model describing a perovskite-based exciton-polariton system. Interestingly, in the case where there is no out-of-plane effective field, part of the non-Hermitian corrections to the group velocity can be written in terms of the diagonal components of the LR quantum metric. This is similar to previous studies of Hermitian systems~\cite{leblanc2021}, which describe the wavepacket COM motion in terms of the quantum metric, and in the so-called exotic zitterbewegung-like motion~\cite{ye2022} described by the LR interband Berry connection. Our work shows the rich novel dynamical effects in non-Hermitian systems, and the deeper connection between dynamical effects and quantum geometry remains an interesting open question for further investigation.

\begin{acknowledgments}
We acknowledge support from the Australian Research Council (ARC) through the Centre of Excellence Grant CE170100039 and the Discovery Early Career Researcher Award DE220100712, and Australian Government Research Training Program (RTP) Scholarship.
\end{acknowledgments}

\section*{Data Availability Statement}
The data that support the findings of this study are available from the corresponding author upon reasonable request.

\section*{Supplementary Materials}
The supplementary materials include the values of parameters used in the simulation, the derivation of the semi-classical equations of motion in Eqs.~(\ref{eq: ZBW},\ref{eq: Hz=0 ZBW}) the main text, which are our main results in this work, as well as the pseudospin dynamics in the non-Hermitian Dirac model.

%aipnum4-2.bst 2019-01-14 (MD) hand-edited version of apsrev4-1.bst
%Control: key (0)
%Control: author (8) initials jnrlst
%Control: editor formatted (1) identically to author
%Control: production of article title (0) allowed
%Control: page (1) range
%Control: year (1) truncated
%Control: production of eprint (0) enabled
%

\end{document}

\preprint{APS/123-QED}

\title{Supplementary Material - Zitterbewegung Effect and Quantum Geometry in Non-Hermitian Exciton-Polariton Systems}% Force line breaks with \\

\author{Y.-M. Robin Hu}
\affiliation{%
Department of Quantum Science and Technology, Research School of Physics, The Australian National University, Canberra, ACT 2601 Australia
}%
\author{Elena A. Ostrovskaya}%
\affiliation{%
Department of Quantum Science and Technology, Research School of Physics, The Australian National University, Canberra, ACT 2601 Australia
}%
\author{Eliezer Estrecho}
\affiliation{%
Department of Quantum Science and Technology, Research School of Physics, The Australian National University, Canberra, ACT 2601 Australia
}%

\maketitle

\section{Values of Parameters}
In this section, we will present the values of the parameters used for the simulation of wavepacket dynamics in the main text. For the exciton-polariton model, we choose the parameters to be $\gamma_2=0$ $\mu$m$^2$eV, $\gamma_4=3.75\times10^{-4}$ $\mu$m$^4$eV, and choose the mean energy, linewidth and effective polariton mass to be $E_0-i\gamma_0=2.306-4.5\times10^{-4}i$ eV and $\hbar^2/2m\approx2.3\times10^{-3}$ $\mu$m$^2$eV. The X-Y splitting parameters are $\alpha=8\times10^{-3}$ eV and $a=10^{-5}$, while the TE-TM splitting parameters are $\beta=10^{-3}$ $\mu$m$^{2}$eV and $b=7.5\times10^{-4}$ $\mu$m$^{2}$eV. This set of parameters ensures that the exceptional points and Fermi arcs lie within low momenta. It also ensures that the two exceptional points in a pair are well separated, allowing us to probe the dynamics on a single exceptional point as well as on the bulk Fermi arcs.

\section{Derivation of Equation of Motion}
In this section, we will present our derivation of the semi-classical equation of motion describing the non-Hermitian zitterbewegung effects in the main text. First, we start with a general non-Hermitian Hamiltonian
\begin{equation}
    \begin{split}
        \hat{H}(\mathbf{k})&=H_0(\mathbf{k})\mathbf{I}+\overrightarrow{\mathbf{H}}(\mathbf{k})\cdot\overrightarrow{\boldsymbol{\sigma}}\\
        \overrightarrow{\mathbf{H}}(\mathbf{k})&=[H_x(\mathbf{k}),H_y(\mathbf{k}),H_z(\mathbf{k})]\\
        \overrightarrow{\boldsymbol{\sigma}}&=[\sigma_x,\sigma_y,\sigma_z]
    \end{split}
\end{equation}
where $H_j=G_j+i\Gamma_j$ for $j=0,x,y,z$. The dynamics of the centre-of-mass position is described by a generalized Ehrenfest theorem
\begin{equation}\label{eq: ehrenfest}
\begin{split}
    \frac{d}{dt}\langle\hat{\mathbf{r}}\rangle&=\frac{i}{\hbar}\langle\hat{H}^\dagger\hat{\mathbf{r}}-\hat{\mathbf{r}}\hat{H}\rangle-\frac{i}{\hbar}\langle\hat{H}^\dagger-\hat{H}\rangle\langle\hat{\mathbf{r}}\rangle\\
    &=\frac{i}{\hbar}\langle[\hat{G},\hat{\mathbf{r}}]\rangle+\frac{1}{\hbar}\langle\{\hat{\Gamma},\hat{\mathbf{r}}\}\rangle-\frac{2}{\hbar}\langle\hat{\Gamma}\rangle\langle\hat{\mathbf{r}}\rangle
\end{split}
\end{equation}
where $[.,.]$ denotes commutator and $\{.,.\}$ denotes anti-commutator, and $\langle\hat{A}\rangle=\int_\mathbf{k}\langle\psi|\hat{A}|\psi\rangle/\int_\mathbf{k}\langle\psi|\psi\rangle$ for any operator $\hat{A}$.

We assume the wavepacket to take the form of $|W\rangle=\int_\mathbf{k}w_\mathbf{k}e^{i\mathbf{k}\cdot\mathbf{r}}|\psi\rangle$ where $w_\mathbf{k}$ is a Gaussian and $|\psi\rangle$ is the spinor wavefunction. Each term can then be solved individually.

We assume that the Hamiltonian has no spatial dependence, and only dependence on the momentum. The first term in Eq.~(\ref{eq: ehrenfest}) can simply be written as
\begin{equation}
    \frac{i}{\hbar}\langle[\hat{G},\hat{\mathbf{r}}]\rangle=\frac{1}{\hbar}\langle\partial_\mathbf{k}\hat{G}\rangle.
\end{equation}

The last term in Eq.~(\ref{eq: ehrenfest}), $\langle\hat{\Gamma}\rangle\langle\hat{\mathbf{r}}\rangle$, can be expanded as
\begin{equation}
    \begin{split}
\langle\hat{\Gamma}\rangle\langle\hat{\mathbf{r}}\rangle&=\frac{1}{N^2}\Big(\int_\mathbf{k}|w_\mathbf{k}|^2\langle\psi|\hat{\Gamma}|\psi\rangle\Big)\Big(\int_\mathbf{k}w_\mathbf{k}^*\langle\psi|i\partial_\mathbf{k}w_\mathbf{k}|\psi\rangle\Big)\\
        &=\frac{1}{N^2}\Big(\int_\mathbf{k}|w_\mathbf{k}|^2\langle\psi|\hat{\Gamma}|\psi\rangle\Big)\Big(\int_\mathbf{k}|w_\mathbf{k}|^2\langle\psi|i\partial_\mathbf{k}\psi\rangle+\int_\mathbf{k}(w_\mathbf{k}^*i\partial_\mathbf{k}w_\mathbf{k})\langle\psi|\psi\rangle\Big),
    \end{split}
\end{equation}
where $N=\int_\mathbf{k}|w_\mathbf{k}|^2\langle\psi|\psi\rangle$ is the norm, which varies with time due to the growth or loss. We can then separate the amplitude and phase of the Gaussian wavefunction as $w_\mathbf{k}=|w_\mathbf{k}|e^{i\varphi}$, then the last term can be rewritten as
\begin{equation}
    \begin{split}
        \int_\mathbf{k}(w_\mathbf{k}^*i\partial_\mathbf{k}w_\mathbf{k})\langle\psi|\psi\rangle&=\int_\mathbf{k}|w_\mathbf{k}|i\partial_\mathbf{k}|w_\mathbf{k}|\langle\psi|\psi\rangle-\int_\mathbf{k}\partial_\mathbf{k}\varphi|w_\mathbf{k}|^2\langle\psi|\psi\rangle\\
        &=-\frac{i}{2}\int_\mathbf{k}|w_\mathbf{k}|^2\partial_\mathbf{k}(\langle\psi|\psi\rangle)-\int_\mathbf{k}\partial_\mathbf{k}\varphi|w_\mathbf{k}|^2\langle\psi|\psi\rangle
    \end{split}
\end{equation}
where we employ integral by parts to obtain the first term in the second line.

After plugging back in $\langle\hat{\Gamma}\rangle\langle\hat{\mathbf{r}}\rangle$, it can be written in the form
\begin{equation}
    \begin{split}
        \langle\hat{\Gamma}\rangle\langle\hat{\mathbf{r}}\rangle&=\frac{1}{N^2}\Big(\int_\mathbf{k}|w_\mathbf{k}|^2\langle\psi|\hat{\Gamma}|\psi\rangle\Big)\Big(\int_\mathbf{k}|w_\mathbf{k}|^2\langle\psi|i\partial_\mathbf{k}\psi\rangle-\frac{i}{2}\int_\mathbf{k}|w_\mathbf{k}|^2\partial_\mathbf{k}(\langle\psi|\psi\rangle)-\int_\mathbf{k}\partial_\mathbf{k}\varphi|w_\mathbf{k}|^2\langle\psi|\psi\rangle\Big).
    \end{split}
\end{equation}
Assuming that the wavepacket is narrow in momentum space, $|w_\mathbf{k}|^2\sim \delta(\mathbf{k}-\mathbf{k}_c)$, the term $\langle\hat{\Gamma}\rangle\langle\hat{\mathbf{r}}\rangle$ can finally be written as
\begin{equation}
    \langle\hat{\Gamma}\rangle\langle\hat{\mathbf{r}}\rangle=\frac{\langle\psi|\hat{\Gamma}|\psi\rangle}{\langle\psi|\psi\rangle}\Big(\frac{i}{2}\frac{\langle\psi|\partial_\mathbf{k}\psi\rangle}{\langle\psi|\psi\rangle}-\frac{i}{2}\frac{\langle\partial_\mathbf{k}\psi|\psi\rangle}{\langle\psi|\psi\rangle}-\partial_\mathbf{k}\varphi\Big)\bigg|_{\mathbf{k}=\mathbf{k}_c}.
\end{equation}
Similarly, the term $\langle\hat{\mathbf{r}}\hat{\Gamma}\rangle$ can be simplified as
\begin{equation}
    \begin{split}
        \langle\hat{\mathbf{r}}\hat{\Gamma}\rangle=&\frac{1}{N}\int_\mathbf{k}w_\mathbf{k}^*\langle\psi|i\partial_\mathbf{k}\hat{\Gamma}w_\mathbf{k}|\psi\rangle\\
        =&\frac{1}{N}\Big(\int_\mathbf{k}|w_\mathbf{k}|^2\langle\psi|(i\partial_\mathbf{k}\hat{\Gamma})|\psi\rangle+\int_\mathbf{k}|w_\mathbf{k}|^2\langle\psi|\hat{\Gamma}|i\partial_\mathbf{k}\psi\rangle+\int_\mathbf{k}(w_\mathbf{k}^*i\partial_\mathbf{k}w_\mathbf{k})\langle\psi|\hat{\Gamma}|\psi\rangle\Big)\\
        =&\frac{1}{N}\int_\mathbf{k}|w_\mathbf{k}|^2\Big(\langle\psi|(i\partial_\mathbf{k}\hat{\Gamma})|\psi\rangle+\langle\psi|\hat{\Gamma}|i\partial_\mathbf{k}\psi\rangle-\partial_\mathbf{k}\varphi\langle\psi|\hat{\Gamma}|\psi\rangle-\frac{i}{2}\partial_\mathbf{k}(\langle\psi|\hat{\Gamma}|\psi\rangle)\Big)\\
        =&\Big(\frac{\langle\psi|(i\partial_\mathbf{k}\hat{\Gamma})|\psi\rangle}{\langle\psi|\psi\rangle}+\frac{\langle\psi|\hat{\Gamma}|i\partial_\mathbf{k}\psi\rangle}{\langle\psi|\psi\rangle}-\partial_\mathbf{k}\varphi\frac{\langle\psi|\hat{\Gamma}|\psi\rangle}{\langle\psi|\psi\rangle}-\frac{i}{2}\frac{\partial_\mathbf{k}(\langle\psi|\hat{\Gamma}|\psi\rangle)}{\langle\psi|\psi\rangle}\Big)\bigg|_{\mathbf{k}=\mathbf{k}_c}\\
        =&\Big(\frac{i}{2}\frac{\langle\psi|(\partial_\mathbf{k}\hat{\Gamma})|\psi\rangle}{\langle\psi|\psi\rangle}+\frac{i}{2}\frac{\langle\psi|\hat{\Gamma}|\partial_\mathbf{k}\psi\rangle}{\langle\psi|\psi\rangle}-\frac{i}{2}\frac{\langle\partial_\mathbf{k}\psi|\hat{\Gamma}|\psi\rangle}{\langle\psi|\psi\rangle}-\partial_\mathbf{k}\varphi\frac{\langle\psi|\hat{\Gamma}|\psi\rangle}{\langle\psi|\psi\rangle}\Big)\bigg|_{\mathbf{k}=\mathbf{k}_c}\\
        \langle\hat{\Gamma}\hat{\mathbf{r}}\rangle=&\frac{1}{N}\int_\mathbf{k}w_\mathbf{k}^*\langle\psi|\hat{\Gamma}i\partial_\mathbf{k}w_\mathbf{k}|\psi\rangle\\
        =&\frac{1}{N}\Big(\int_\mathbf{k}|w_\mathbf{k}|^2\langle\psi|\hat{\Gamma}|i\partial_\mathbf{k}\psi\rangle+\int_\mathbf{k}(w_\mathbf{k}^*i\partial_\mathbf{k}w_\mathbf{k})\langle\psi|\hat{\Gamma}|\psi\rangle\Big)\\
        =&\Big(-\frac{i}{2}\frac{\langle\psi|(\partial_\mathbf{k}\hat{\Gamma})|\psi\rangle}{\langle\psi|\psi\rangle}+\frac{i}{2}\frac{\langle\psi|\hat{\Gamma}|\partial_\mathbf{k}\psi\rangle}{\langle\psi|\psi\rangle}-\frac{i}{2}\frac{\langle\partial_\mathbf{k}\psi|\hat{\Gamma}|\psi\rangle}{\langle\psi|\psi\rangle}-\partial_\mathbf{k}\varphi\frac{\langle\psi|\hat{\Gamma}|\psi\rangle}{\langle\psi|\psi\rangle}\Big)\bigg|_{\mathbf{k}=\mathbf{k}_c}.
    \end{split}
\end{equation}

By combining these equations, Eq.~(\ref{eq: ehrenfest}) can be simplified as
\begin{equation}\label{eq: eom general}
    \begin{split}
        \frac{d}{dt}\langle\hat{\mathbf{r}}\rangle=&\frac{1}{\hbar}\langle\partial_\mathbf{k}\hat{G}\rangle+\frac{i}{\hbar}\Big(\frac{\langle\psi|\hat{\Gamma}|\partial_\mathbf{k}\psi\rangle}{\langle\psi|\psi\rangle}-\frac{\langle\partial_\mathbf{k}\psi|\hat{\Gamma}|\psi\rangle}{\langle\psi|\psi\rangle}\Big)-\frac{i}{\hbar}\langle\hat{\Gamma}\rangle\Big(\frac{\langle\psi|\partial_\mathbf{k}\psi\rangle}{\langle\psi|\psi\rangle}-\frac{\langle\partial_\mathbf{k}\psi|\psi\rangle}{\langle\psi|\psi\rangle}\Big)\bigg|_{\mathbf{k}=\mathbf{k}_c}
    \end{split}
\end{equation}
The spinor wavefcuntion can be written as $|\psi\rangle=e^{-\frac{i\hat{H}t}{\hbar}}|\psi_0\rangle$ where $|\psi_0\rangle$ is the initial state. Assuming that $|\psi_0\rangle$ has no $\mathbf{k}$-dependence, $\langle\partial_\mathbf{k}\psi|$ and $|\partial_\mathbf{k}\psi\rangle$ can be rewritten as
\begin{equation}\label{eq: dkpsi}
    \begin{split}
        \langle\partial_\mathbf{k}\psi|&=\langle\psi_0|\partial_\mathbf{k}e^{\frac{i\hat{H}^\dagger t}{\hbar}}\\
        |\partial_\mathbf{k}\psi\rangle&=\partial_\mathbf{k}e^{-\frac{i\hat{H}t}{\hbar}}|\psi_0\rangle.
    \end{split}
\end{equation}
The first line in Eqs.~(\ref{eq: dkpsi}) can be solved using the identity of the derivative of the exponential map \cite{tuynman1995,rossman2002}
\begin{equation}\label{eq: dk exp}
    \begin{split}
        \partial_\mathbf{k}e^{\frac{i\hat{H}^\dagger t}{\hbar}}&=e^{\frac{i\hat{H}^\dagger t}{\hbar}}\bigg(\frac{1-exp\Big(-\text{ad}_{\frac{i\hat{H}^\dagger t}{\hbar}}\Big)}{\text{ad}_{\frac{i\hat{H}^\dagger t}{\hbar}}}\partial_\mathbf{k}\Big(\frac{i\hat{H}^\dagger t}{\hbar}\Big)\bigg)\\
        &=e^{\frac{i\hat{H}^\dagger t}{\hbar}}\sum_{n=0}^\infty\frac{(-1)^n}{(n+1)!}\Big(\frac{it}{\hbar}\Big)^{n+1}\text{ad}^n_{\hat{H}^\dagger}\Big(\partial_\mathbf{k}\hat{H}^\dagger\Big)
    \end{split}
\end{equation}
where for matrices $X,Y$ $\text{ad}_X$ denotes the adjoint representation, $\text{ad}_XY=[X,Y]$.

Recall that $\hat{H}^\dagger=H_0^*\mathbf{I}+H_x^*\sigma_x+H_y^*\sigma_y+H_z^* \sigma_z$. The zero-th order, the odd and even order terms take the forms of
\begin{equation}\label{eq: ad series}
    \begin{split}
    \text{ad}_{\hat{H}^\dagger}^0\partial_\mathbf{k}\hat{H}^\dagger&=\partial_\mathbf{k}H_0^*\mathbf{I}+\partial_\mathbf{k}H_x^*\sigma_x+\partial_\mathbf{k}H_y^*\sigma_y+\partial_\mathbf{k}H_z^* \sigma_z\\
    \text{ad}_{\hat{H}^\dagger}^{2n-1}\partial_\mathbf{k}\hat{H}^\dagger&=2^{2n-1}i(H_x^{*2}+H_y^{*2}+H_z^{*2})^{n-1}(A_x^*\sigma_x+A_y^*\sigma_y+A_z^*\sigma_z)\\
    \text{ad}_{\hat{H}^\dagger}^{2n}\partial_\mathbf{k}\hat{H}^\dagger&=2^{2n}(H_x^{*2}+H_y^{*2}+H_z^{*2})^{n-1}(B_x^*\sigma_x+B_y^*\sigma_y+B_z^*\sigma_z)
    \end{split}
\end{equation}
for $n=1,...\infty$. Where $A_i, B_i$ denote
\begin{equation}
    \begin{split}
        A_x&=H_y\partial_\mathbf{k}H_z-H_z\partial_\mathbf{k}H_y\\
        A_y&=H_z\partial_\mathbf{k}H_x-H_x\partial_\mathbf{k}H_z\\
        A_z&=H_x\partial_\mathbf{k}H_y-H_y\partial_\mathbf{k}H_x\\
        B_x&=\partial_\mathbf{k}H_x(H_y^2+H_z^2)-H_x(H_y\partial_\mathbf{k}H_y+H_z\partial_\mathbf{k}H_z)\\
        B_y&=\partial_\mathbf{k}H_y(H_x^2+H_z^2)-H_y(H_x\partial_\mathbf{k}H_x+H_z\partial_\mathbf{k}H_z)\\
        B_z&=\partial_\mathbf{k}H_z(H_x^2+H_y^2)-H_z(H_x\partial_\mathbf{k}H_x+H_y\partial_\mathbf{k}H_y)
    \end{split}
\end{equation}
to simplify the expression. Using these identities and noting that the mean-subtracted eigenenergy and its complex conjugate take the forms of $E=\sqrt{H_x^2+H_y^2+H_z^2}$, $E^*=\sqrt{H_x^{*2}+H_y^{*2}+H_z^{*2}}$, Eq.~(\ref{eq: dk exp}) can then be rewritten as
\begin{equation}\label{eq: dk T dagger}
    \begin{split}
        \partial_\mathbf{k}e^{\frac{i\hat{H}^\dagger t}{\hbar}}=&e^{\frac{i\hat{H}^\dagger t}{\hbar}}\Bigg(\frac{it}{\hbar}\partial_\mathbf{k}H_0^*\mathbf{I}+\frac{it}{\hbar}(\partial_\mathbf{k}H_x^*\sigma_x+\partial_\mathbf{k}H_y^*\sigma_y+\partial_\mathbf{k}H_z^*\sigma_z)\\
        &+\sum_{n=1}^\infty\frac{(-1)^{2n-1}}{(2n)!}\Big(\frac{it}{\hbar}\Big)^{2n}2^{2n-1}i(H_x^{*2}+H_y^{*2}+H_z^{*2})^{n-1}(A_x^*\sigma_x+A_y^*\sigma_y+A_z^*\sigma_z)\\
        &+\sum_{n=1}^\infty\frac{(-1)^{2n}}{(2n+1)!}\Big(\frac{it}{\hbar}\Big)^{2n+1}2^{2n}(H_x^{*2}+H_y^{*2}+H_z^{*2})^{n-1}(B_x^*\sigma_x+B_y^*\sigma_y+B_z^*\sigma_z)\Bigg)\\
        =&e^{\frac{i\hat{H}^\dagger t}{\hbar}}\Bigg(\frac{it}{\hbar}\partial_\mathbf{k}H_0^*\mathbf{I}+\frac{it}{\hbar}(\partial_\mathbf{k}H_x^*\sigma_x+\partial_\mathbf{k}H_y^*\sigma_y+\partial_\mathbf{k}H_z^*\sigma_z)\\
        &-\sum_{n=1}^\infty\frac{i^{2n+1}}{(2n)!}\Big(\frac{t}{\hbar}\Big)^{2n}2^{2n-1}E^{*2n-2}(A_x^*\sigma_x+A_y^*\sigma_y+A_z^*\sigma_z)\\
        &+\sum_{n=1}^\infty\frac{i^{2n+1}}{(2n+1)!}\Big(\frac{t}{\hbar}\Big)^{2n+1}2^{2n}E^{* 2n-2}(B_x^*\sigma_x+B_y^*\sigma_y+B_z^*\sigma_z)\Bigg)\\
        =&e^{\frac{i\hat{H}^\dagger t}{\hbar}}\Bigg(\frac{it}{\hbar}\partial_\mathbf{k}H_0^*\mathbf{I}+\frac{it}{\hbar}(\partial_\mathbf{k}H_x^*\sigma_x+\partial_\mathbf{k}H_y^*\sigma_y+\partial_\mathbf{k}H_z^*\sigma_z)\\
        &-\sum_{n=1}^\infty\frac{i}{2E^{*2}}\frac{(-1)^n}{(2n)!}\Big(\frac{2E^*t}{\hbar}\Big)^{2n}(A_x^*\sigma_x+A_y^*\sigma_y+A_z^*\sigma_z)\\
        &+\sum_{n=1}^\infty\frac{i}{2E^{*3}}\frac{(-1)^n}{(2n+1)!}\Big(\frac{2E^*t}{\hbar}\Big)^{2n+1}(B_x^*\sigma_x+B_y^*\sigma_y+B_z^*\sigma_z)\Bigg)\\
        =&e^{\frac{i\hat{H}^\dagger t}{\hbar}}\Bigg(\frac{it}{\hbar}\partial_\mathbf{k}H_0^*\mathbf{I}+\frac{it}{\hbar}(\partial_\mathbf{k}H_x^*\sigma_x+\partial_\mathbf{k}H_y^*\sigma_y+\partial_\mathbf{k}H_z^*\sigma_z)\\
        &-\frac{i}{2E^{*2}}\Big(\cos{\frac{2E^*t}{\hbar}}-1\Big)(A_x^*\sigma_x+A_y^*\sigma_y+A_z^*\sigma_z)\\
        &+\frac{i}{2E^{*3}}\Big(\sin{\frac{2E^*t}{\hbar}}-\frac{2E^*t}{\hbar}\Big)(B_x^*\sigma_x+B_y^*\sigma_y+B_z^*\sigma_z)\Bigg).
    \end{split}
\end{equation}
Similarly, we can solve $\partial_\mathbf{k}e^{-\frac{i\hat{\mathbf{H}}t}{\hbar}}$ by taking the Hermitian conjugate of the results in Eq.~(\ref{eq: dk T dagger})
\begin{equation}\label{eq: dk T}
    \begin{split}
        \partial_\mathbf{k}e^{-\frac{i\hat{H} t}{\hbar}}=&\Bigg(-\frac{it}{\hbar}\partial_\mathbf{k}H_0\mathbf{I}-\frac{it}{\hbar}(\partial_\mathbf{k}H_x\sigma_x+\partial_\mathbf{k}H_y\sigma_y+\partial_\mathbf{k}H_z\sigma_z)\\
        &+\frac{i}{2E^{2}}\Big(\cos{\frac{2Et}{\hbar}}-1\Big)(A_x\sigma_x+A_y\sigma_y+A_z\sigma_z)\\
        &-\frac{i}{2E^{3}}\Big(\sin{\frac{2Et}{\hbar}}-\frac{2Et}{\hbar}\Big)(B_x\sigma_x+B_y\sigma_y+B_z\sigma_z)\Bigg)e^{-\frac{i\hat{H}t}{\hbar}}.
    \end{split}
\end{equation}
We rewrite the expressions in Eqs.~(\ref{eq: dk T dagger},\ref{eq: dk T}) by defining a new operator $\hat{L}$ as
\begin{equation}
    \hat{L}=\frac{t}{\hbar}\partial_\mathbf{k}H_0\mathbf{I}+\sum_{j=x,y,z}\Bigg(\bigg(\frac{t}{\hbar}\partial_\mathbf{k}H_j-\frac{\cos{\frac{2Et}{\hbar}}-1}{2E^{2}}A_j+\frac{\sin{\frac{2Et}{\hbar}}-\frac{2Et}{\hbar}}{2E^{3}}B_j\bigg)\sigma_j\Bigg)
\end{equation}
and therefore the derivatives of the time evolution operator and its conjugate take the forms of
\begin{equation}
    \begin{split}
        \partial_\mathbf{k}e^{-\frac{i\hat{H} t}{\hbar}}=&-i\hat{L}e^{-\frac{i\hat{H} t}{\hbar}}\\
        \partial_\mathbf{k}e^{\frac{i\hat{H}^\dagger t}{\hbar}}=&ie^{\frac{i\hat{H}^\dagger t}{\hbar}}\hat{L}^\dagger,
    \end{split}
\end{equation}
and the equation of motion in Eq.~(\ref{eq: eom general}) can be rewritten as
\begin{equation}
    \frac{d}{dt}\langle\hat{\mathbf{r}}\rangle=\frac{1}{\hbar}\langle\partial_\mathbf{k}\hat{G}\rangle+\frac{1}{\hbar}\langle\hat{\Gamma}\hat{L}+\hat{L}^\dagger\hat{\Gamma}\rangle-\frac{1}{\hbar}\langle\hat{\Gamma}\rangle\langle\hat{L}+\hat{L}^\dagger\rangle\bigg|_{\mathbf{k}=\mathbf{k}_c}.
\end{equation}
Since the coefficients in $\hat{L}$ can all be complex-valued, we can further simplify the equation by defining two new operators corresponding to the components corresponding to the real   and imaginary parts of the coefficients in $\hat{L}$
\begin{equation}\label{eq: re im L}
    \begin{split}
        \hat{L}^{Re}&=\frac{t}{\hbar}\partial_\mathbf{k}G_0\mathbf{I}+\sum_{j=x,y,z}\Bigg(\frac{t}{\hbar}\partial_\mathbf{k}G_j\sigma_j+\operatorname{Re}\bigg[-\frac{\cos{\frac{2Et}{\hbar}}-1}{2E^{2}}A_j+\frac{\sin{\frac{2Et}{\hbar}}-\frac{2Et}{\hbar}}{2E^{3}}B_j\bigg]\sigma_j\Bigg)\\
        \hat{L}^{Im}&=\frac{t}{\hbar}\partial_\mathbf{k}\Gamma_0\mathbf{I}+\sum_{j=x,y,z}\Bigg(\frac{t}{\hbar}\partial_\mathbf{k}\Gamma_j\sigma_j+\operatorname{Im}\bigg[-\frac{\cos{\frac{2Et}{\hbar}}-1}{2E^{2}}A_j+\frac{\sin{\frac{2Et}{\hbar}}-\frac{2Et}{\hbar}}{2E^{3}}B_j\bigg]\sigma_j\Bigg)
    \end{split}
\end{equation}
such that $\hat{L}=\hat{L}^{Re}+i\hat{L}^{Im}$ and $\hat{L}^\dagger=\hat{L}^{Re}-i\hat{L}^{Im}$.

The equation of motion in Eq.~(\ref{eq: eom general}) can then be further simplified as
\begin{equation}\label{eq: suppl ZBW}
    \frac{d}{dt}\langle\hat{\mathbf{r}}\rangle=\frac{1}{\hbar}\langle\partial_\mathbf{k}\hat{G}\rangle+\frac{1}{\hbar}\langle\{\hat{\Gamma},\hat{L}^{Re}\}\rangle+\frac{i}{\hbar}\langle[\hat{\Gamma},\hat{L}^{Im}]\rangle-\frac{2}{\hbar}\langle\hat{\Gamma}\rangle\langle\hat{L}^{Re}\rangle\bigg|_{\mathbf{k}=\mathbf{k}_c}.
\end{equation}
We can then simplify the equation of motion term by term to obtain the main result in Eq.~(\ref{eq: ZBW}) in the main text. The first term in Eq.~(\ref{eq: suppl ZBW}) also appears in the Hermitian case and takes the form of
\begin{equation}
    \frac{1}{\hbar}\langle\partial_\mathbf{k}\hat{G}\rangle=\frac{1}{\hbar}\Big(\partial_\mathbf{k}G_0+\partial_\mathbf{k}G_xS_x+\partial_\mathbf{k}G_yS_y+\partial_\mathbf{k}G_zS_z\Big)
\end{equation}
where $S_j=\langle\sigma_j\rangle$ for $j=x,y,z$. The second term in Eq.~(\ref{eq: suppl ZBW}) can solved using the anti-commutation relations of the Pauli matrices and the identity matrix $\{\mathbf{I},\sigma_j\}=2\sigma_j$, $\{\sigma_i,\sigma_j\}=2\delta_{ij}\mathbf{I}$
\begin{equation}\label{eq: suppl ZBW 2nd term}
\begin{split}
    \frac{1}{\hbar}\langle\{\hat{\Gamma},\hat{L}^{Re}\}\rangle=&\frac{1}{\hbar}\sum_{i,j=x,y,z}\bigg(\langle\{\Gamma_0\mathbf{I},\partial_\mathbf{k}G_0\mathbf{I}+\frac{t}{\hbar}\partial_\mathbf{k}G_j\sigma_j+\operatorname{Re}\bigg[-\frac{\cos{\frac{2Et}{\hbar}}-1}{2E^{2}}A_j+\frac{\sin{\frac{2Et}{\hbar}}-\frac{2Et}{\hbar}}{2E^{3}}B_j\bigg]\sigma_j\}\rangle\\
    &+\langle\{\Gamma_i\sigma_i,\partial_\mathbf{k}G_0\mathbf{I}+\frac{t}{\hbar}\partial_\mathbf{k}G_j\sigma_j+\operatorname{Re}\bigg[-\frac{\cos{\frac{2Et}{\hbar}}-1}{2E^{2}}A_j+\frac{\sin{\frac{2Et}{\hbar}}-\frac{2Et}{\hbar}}{2E^{3}}B_j\bigg]\sigma_j\}\rangle\bigg)\\
    =&\frac{2}{\hbar}\sum_{j=x,y,z}\bigg(\Gamma_0\partial_\mathbf{k}G_0+\frac{t}{\hbar}\Gamma_0\partial_\mathbf{k}G_jS_j+\operatorname{Re}\bigg[-\frac{\cos{\frac{2Et}{\hbar}}-1}{2E^{2}}\Gamma_0A_j+\frac{\sin{\frac{2Et}{\hbar}}-\frac{2Et}{\hbar}}{2E^{3}}\Gamma_0B_j\bigg]S_j\\
    &+\Gamma_j\partial_\mathbf{k}G_0S_j+\frac{t}{\hbar}\Gamma_j\partial_\mathbf{k}G_j+\operatorname{Re}\bigg[-\frac{\cos{\frac{2Et}{\hbar}}-1}{2E^{2}}A_j+\frac{\sin{\frac{2Et}{\hbar}}-\frac{2Et}{\hbar}}{2E^{3}}B_j\bigg]\Gamma_j\bigg),
\end{split}
\end{equation}
and the third term in Eq.~(\ref{eq: suppl ZBW}) can be rewritten using the commutation relation $[\sigma_i,\sigma_j]=2i\epsilon_{ijk}\sigma_k$
\begin{equation}
    \begin{split}
        \frac{i}{\hbar}\langle[\hat{\Gamma},\hat{L}^{Im}]\rangle=&-\frac{2}{\hbar}\sum_{i,j=x,y,z}\epsilon_{ijk}\bigg(\frac{t}{\hbar}\Gamma_i\partial_\mathbf{k}\Gamma_j+\Gamma_i\operatorname{Im}\bigg[-\frac{\cos{\frac{2Et}{\hbar}}-1}{2E^{2}}A_j+\frac{\sin{\frac{2Et}{\hbar}}-\frac{2Et}{\hbar}}{2E^{3}}B_j\bigg]\bigg)S_k.
    \end{split}
\end{equation}
Finally, the last term in Eq.~(\ref{eq: suppl ZBW}) is simply
\begin{equation}\label{eq: suppl ZBW 4th term}
    \begin{split}
        -\frac{2}{\hbar}\langle\hat{\Gamma}\rangle\langle\hat{L}^{Re}\rangle=&-\frac{2}{\hbar}\bigg(\Gamma_0+\sum_{i=x,y,z}\Gamma_i S_i\bigg)\bigg(\frac{t}{\hbar}\partial_\mathbf{k}G_0+\sum_{j=x,y,z}\bigg(\frac{t}{\hbar}\partial_\mathbf{k}G_jS_j\\
        &+\operatorname{Re}\bigg[-\frac{\cos{\frac{2Et}{\hbar}}-1}{2E^{2}}A_j+\frac{\sin{\frac{2Et}{\hbar}}-\frac{2Et}{\hbar}}{2E^{3}}B_j\bigg]S_j\bigg)\bigg)\\
        =&-\frac{2}{\hbar}\Gamma_0\Bigg(\frac{t}{\hbar}\partial_\mathbf{k}G_0+\sum_{j=x,y,z}\bigg(\frac{t}{\hbar}\partial_\mathbf{k}G_jS_j+\operatorname{Re}\bigg[-\frac{\cos{\frac{2Et}{\hbar}}-1}{2E^{2}}A_j+\frac{\sin{\frac{2Et}{\hbar}}-\frac{2Et}{\hbar}}{2E^{3}}B_j\bigg]S_j\bigg)\Bigg)\\
        &-\frac{2t}{\hbar^2}\partial_\mathbf{k}G_0\bigg(\Gamma_0+\sum_{i=x,y,z}\Gamma_i S_i\bigg)\\
        &-\frac{2}{\hbar}\Big(\sum_{i=x,y,z}\Gamma_i S_i\Big)\bigg(\sum_{j=x,y,z}\frac{t}{\hbar}\partial_\mathbf{k}G_jS_j+\operatorname{Re}\bigg[-\frac{\cos{\frac{2Et}{\hbar}}-1}{2E^{2}}A_j+\frac{\sin{\frac{2Et}{\hbar}}-\frac{2Et}{\hbar}}{2E^{3}}B_j\bigg]S_j\bigg)
    \end{split}
\end{equation}
where we note that the first two lines in the final form of Eq.~(\ref{eq: suppl ZBW 4th term}) will be cancelled out with terms in Eq.~(\ref{eq: suppl ZBW 2nd term}) and only the terms in $O(S^2)$ will remain. 

After summing up and rearranging the terms, Eq.~(\ref{eq: suppl ZBW}) can be rewritten into
\begin{equation}
    \begin{split}
        \Dot{\mathbf{r}}=&\frac{1}{\hbar}\partial_\mathbf{k}G_0+\frac{1}{\hbar}\sum_i(\partial_\mathbf{k}G_i)S_i+\frac{2t}{\hbar^2}\sum_i\Gamma_i\partial_\mathbf{k}G_i-\frac{2t}{\hbar^2}\sum_{i,j}\epsilon_{ijk}\Gamma_i(\partial_\mathbf{k}\Gamma_j)S_k\\
        &-\frac{2t}{\hbar^2}\Big(\sum_i\Gamma_i S_i\Big)\Big(\sum_j(\partial_\mathbf{k}G_j)S_j\Big)+\frac{2}{\hbar}\sum_i\operatorname{Re}\Big[-\frac{\cos{\frac{2Et}{\hbar}}-1}{2E^2}\Gamma_i A_i+\frac{\sin{\frac{2Et}{\hbar}}-\frac{2Et}{\hbar}}{2E^3}\Gamma_i B_i\Big]\\
        &-\frac{2}{\hbar}\sum_{i,j}\epsilon_{ijk}\operatorname{Im}\Big[-\frac{\cos{\frac{2Et}{\hbar}}-1}{2E^2}\Gamma_i A_j S_k+\frac{\sin{\frac{2Et}{\hbar}}-\frac{2Et}{\hbar}}{2E^3}\Gamma_i B_j S_k\Big]\\
        &-\frac{2}{\hbar}(\sum_i\Gamma_i S_i)\bigg(\sum_j\operatorname{Re}\Big[-\frac{\cos{\frac{2Et}{\hbar}}-1}{2E^2}A_j S_j+\frac{\sin{\frac{2Et}{\hbar}}-\frac{2Et}{\hbar}}{2E^3}B_j S_j\Big]\bigg).
    \end{split}
\end{equation}
After rearranging the terms, this is our main result in the main text.

\section{Quantum Metric Tensor in the Equation of Motion}
In the special case with $H_z=0$, the identities of $\text{ad}_{\hat{H}}^n\hat{H}$ now take the forms
\begin{equation}
    \begin{split}
    \text{ad}_{\hat{H}}^0\partial_\mathbf{k}\hat{H}&=\partial_\mathbf{k}H_0\mathbf{I}+\partial_\mathbf{k}H_x\sigma_x+\partial_\mathbf{k}H_y\sigma_y\\
    \text{ad}_{\hat{H}}^{2n-1}\partial_\mathbf{k}\hat{H}&=2^{2n-1}i(H_x^2+H_y^2+H_z^2)^{n-1}(H_x\partial_\mathbf{k}H_y-H_y\partial_\mathbf{k}H_x)\sigma_z\\
    \text{ad}_{\hat{H}}^{2n}\partial_\mathbf{k}\hat{H}&=2^{2n}(H_x^2+H_y^2+H_z^2)^{n-1}(H_x\partial_\mathbf{k}H_y-H_y\partial_\mathbf{k}H_x)(-H_y\sigma_x+H_x\sigma_y).
    \end{split}
\end{equation}
We also note the these terms can be identified with components of the left-right quantum metric tensor (LR QMT), $g_{n,\mu\nu}^{LR}$, which are defined as \cite{zhang2019,zhu2021,brody2013,ye2023}
\begin{equation}
    g_{n,\mu\nu}^{LR}=\frac{1}{2}\Big(\langle\partial_\mu\psi_n^L|\partial_\nu\psi_n^R\rangle+\langle\partial_\nu\psi_n^L|\partial_\mu\psi_n^R\rangle\Big)+\langle\partial_\mu\psi_n^L|\psi_n^R\rangle\langle\psi_n^L|\partial_\nu\psi_n^R\rangle.
\end{equation}
The components of the LR QMT also follow the identity \cite{zhang2019,piechon2016,bleu2018}
\begin{equation}
\begin{split}
    g_{n,\mu\nu}^{LR}&=\frac{1}{2}\sum_{m\neq n}\bigg((\mathbf{A}_{nm}^{LR})_\mu(\mathbf{A}_{mn}^{LR})_\nu+(\mathbf{A}_{mn}^{LR})_\mu(\mathbf{A}_{nm}^{LR})_\nu\bigg)\\
    &=\frac{1}{2}\sum_{m\neq n}\bigg(\frac{\langle\psi_m^L|\partial_\mu\hat{H}|\psi_n^R\rangle\langle\psi_n^L|\partial_\nu\hat{H}|\psi_m^R\rangle}{(E_m-E_n)^2}+\frac{\langle\psi_n^L|\partial_\mu\hat{H}|\psi_m^R\rangle\langle\psi_m^L|\partial_\nu\hat{H}|\psi_n^R\rangle}{(E_m-E_n)^2}\bigg)
\end{split}
\end{equation}
where $\mathbf{A}_{nm}^{LR}=\langle\psi_n^L|i\partial_\mathbf{k}\psi_m^R\rangle$ denotes the inter-band Berry connection.

In a two-band non-Hermitian system, the LR QMT of the two eigenstates coincide $g_{+,\mu\nu}^{LR}=g_{-,\mu\nu}^{LR}=g_{\mu\nu}^{LR}$  and in the special case with $H_z=0$, the eigenstates take the forms of
\begin{equation}
    \begin{split}
        |\psi_\pm^R\rangle&=\frac{1}{\sqrt{2E^2}}\begin{pmatrix}
            \pm E\\ H_x+iH_y
        \end{pmatrix}\\
        \langle\psi_\pm^L|&=\frac{1}{\sqrt{2E^2}}\begin{pmatrix}
            \pm E & H_x-iH_y
        \end{pmatrix}.
    \end{split}
\end{equation}
We note that the diagonal elements of the LR QMT and the inter-band Berry connection take the forms of
\begin{equation}
\begin{split}
    g_{\mathbf{kk}}^{LR}&=\frac{(H_x\partial_\mathbf{k}H_y-H_y\partial_\mathbf{k}H_x)^2}{4E^4}\\
    \mathbf{A}_{\pm\mp}^{LR}&=\mathbf{A}_{\mp\pm}^{LR}=\frac{H_y\partial_\mathbf{k}H_x-H_x\partial_\mathbf{k}H_y}{2E^2}.
\end{split}
\end{equation}
We therefore identify the term $(H_x\partial_\mathbf{k}H_y-H_y\partial_\mathbf{k}H_x)/(2E^2)$ with $\sqrt{g_{\mathbf{kk}}^{LR}}$ up to a sign. After this, the equation of motion in Eq.~(\ref{eq: Hz=0 ZBW}) in the main text can be derived similarly to the derivation presented earlier in this section.

\section{Psuedospin Dynamics in the Non-Hermitian Dirac Model}
\begin{figure}[t]
    \centering
    \includegraphics[width=0.7\linewidth]{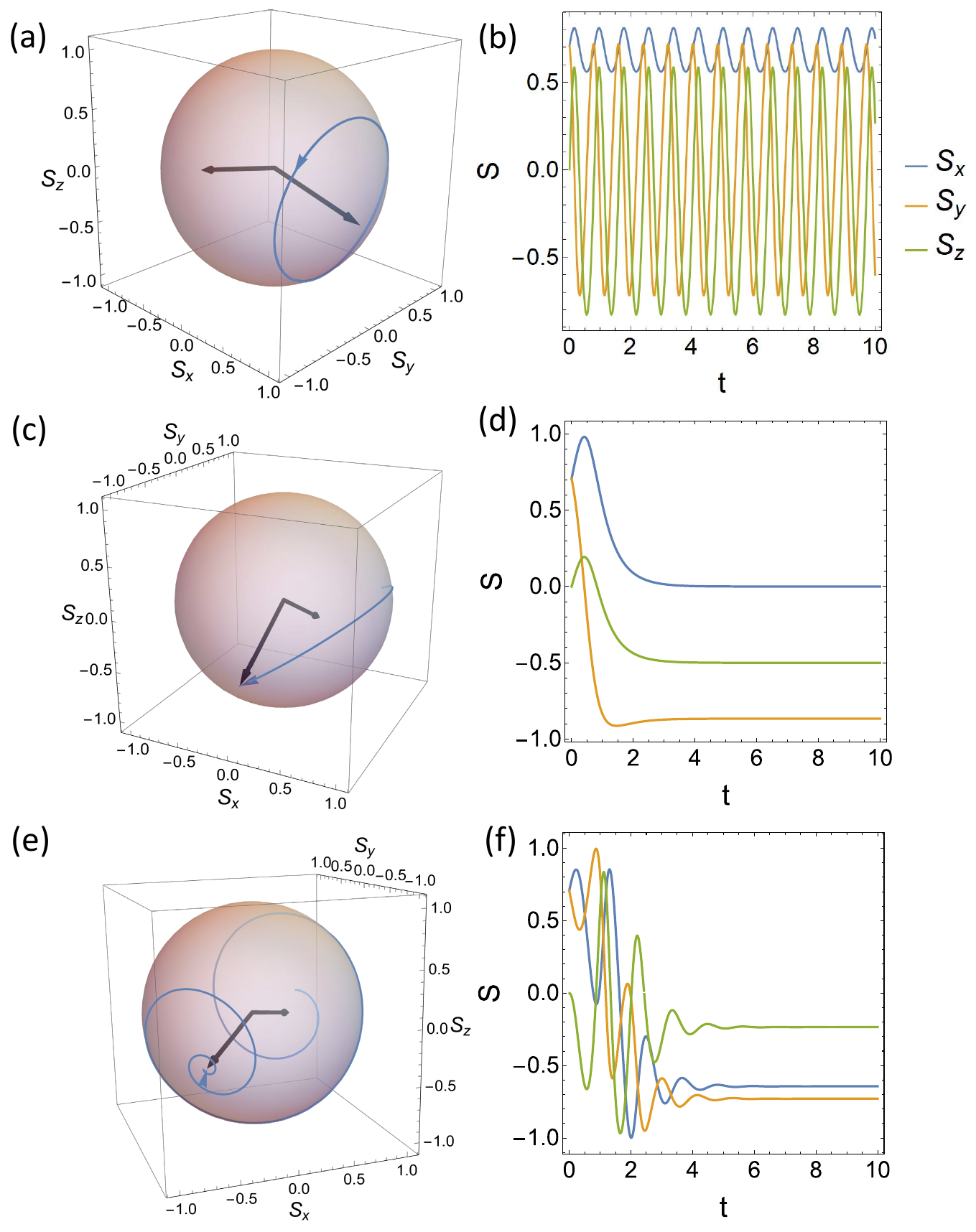}
    \caption{Pseudospin dynamics for a non-Hermitian Dirac model plotted on Bloch spheres and plotted with time. (a,b) are on the imaginary Fermi arc at $\mathbf{k}=(4,0)$,  (c,d)) are on the bulk Fermi arc at $\mathbf{k}=(0.5,0)$ and (e,f) are away from the Fermi arcs at $\mathbf{k}=(2,2)$. The blue arrows represent the trajectories of the pseudospins and the black arrows represent the pseudospins of the two eigenstates. The initial pseudospin configuration is chosen to be $\mathbf{S}=(1/\sqrt{2},1/\sqrt{2},0)$ an example.}
    \label{fig: zbw supp}
\end{figure}

In this section, we present the pseudospin dynamics corresponding to the simulation of the wavepackets in the non-Hermitian Dirac model. The non-Hermitian Dirac model has the Hamiltonian of \cite{solnyshkov2021}
\begin{equation}
    \hat{H}(\mathbf{k})=k_x\sigma_x+(k_y-i\kappa)\sigma_y
\end{equation}
where we set $\kappa=-1$. In Fig.~\ref{fig: zbw supp}, we show the dynamics of the pseudospin in the non-Hermitian Dirac model on the imaginary Fermi arc [see Fig.~\ref{fig: zbw supp}(a,b)], on the bulk Fermi arc [see Fig.~\ref{fig: zbw supp}(c,d)] and away from the Fermi arcs [see Fig.~\ref{fig: zbw supp}(e,f)]. The initial conditions of these three cases correspond to the three cases shown in the simulation of wavepacket dynamics in the same non-Hermitian Dirac model presented in the main text.

